\documentclass[letter,twocolumn]{jpsj3}

\usepackage{txfonts}
\usepackage{graphicx}
\usepackage{dcolumn}
\usepackage{bm}
\usepackage{color}
\usepackage{mathrsfs}

\title{ 
Thermal Einstein--de Haas Effect Induced by Chiral Phonons in Carbon Nanotubes
}

\author{Raimu Akimoto$^{1}$, Hiroyasu Matsuura$^{2}$, and Takahiro Yamamoto$^{1,3,*}$}

\inst{$^1$ Department of Physics, Tokyo University of Science, 1-3 Kagurazaka, Shinjuku, Tokyo 162-8601, Japan \\
$^2$ National Institute of Advanced Industrial Science and Technology (AIST), Tsukuba 305-8568, Ibaraki, Japan \\
$^3$ RIST, Tokyo University of Science, 1-3 Kagurazaka Shinjuku, Tokyo 162-8601, Japan} 

\abst{
We investigate the effects of chirality on phonon thermal transport in semiconducting chiral single-walled carbon nanotubes (SWCNTs) 
using lattice dynamics combined with Boltzmann transport theory. We find that transverse acoustic and optical phonon modes, which 
are degenerate in nonchiral zigzag and armchair SWCNTs, are split in chiral SWCNTs, giving rise to finite phonon angular momentum 
associated with circular motion of individual atoms. This angular momentum is most efficiently generated in small-diameter nanotubes with intermediate 
chiral angles. Consequently, chiral SWCNTs are predicted to undergo thermally induced rigid-body rotation with an experimentally 
observable angular velocity via the thermal Einstein--de Haas effect.
}

\begin{document}
\maketitle


The search for phenomena related to chirality has been actively pursued in various research fields, including the dynamics of chiral magnets~\cite{rf:Togawa}, 
nano-optics~\cite{rf:Kelly}, and transport phenomena such as chiral-induced spin selectivity~\cite{rf:Bloom, rf:Naaman}. In particular, lattice dynamics in chiral crystals 
have recently attracted considerable attention, as phonons in chiral crystals can exhibit unconventional motions in which atomic nuclei undergo rotational 
motion, leading to finite phonon angular momentum~\cite{rf:Zhang01, rf:Zhang02, rf:Kishine, rf:Ishito01, rf:Ohe, rf:Oishi, rf:Ueda, rf:Ishito02, rf:Hamada, 
rf:Tsunetsugu, rf:Chen, rf:Matsuo, rf:H.Zhang}. Such phonon modes are referred to as chiral phonons and have been experimentally observed in a variety of materials, 
including $\alpha$-HgS~\cite{rf:Ishito01}, $\alpha$-quartz~\cite{rf:Ohe, rf:Oishi}, and other chiral crystals~\cite{rf:Ueda, rf:Ishito02}. From the theoretical perspective, 
chiral phonons have been extensively studied using both first-principles calculations~\cite{rf:Hamada} and model analyses~\cite{rf:Tsunetsugu}. Moreover, 
it has been proposed based on both experimental and theoretical studies that the angular momentum carried by chiral phonons can be transferred to the spin 
angular momentum of electrons~\cite{rf:Matsuo}, highlighting chiral phonons as a promising concept in the context of spintronics. Furthermore, theoretical 
studies have suggested that applying a temperature gradient to a chiral crystal can induce macroscopic rotation of the crystal as a consequence of angular 
momentum conservation~\cite{rf:Hamada, rf:Ohe, rf:Chen}. This phenomenon is analogous to the Einstein--de Haas (EdH) effect, in which a system undergoes 
mechanical rotation in response to magnetization~\cite{rf:Einstein}, and is therefore referred to as the thermal EdH effect~\cite{rf:Hamada, rf:Ohe, rf:Chen}. 
Despite extensive theoretical interest, however, the thermal EdH effect has not yet been experimentally observed.

Carbon nanotubes (CNTs), which are experimentally realizable one-dimensional (1D) nanomaterials~\cite{rf:Iijima}, are broadly classified into three distinct structural 
types (zigzag, armchair, and chiral) depending on their rolling geometry~\cite{rf:Saito01, rf:Mintmire}. The former two types, namely zigzag and armchair CNTs, 
are achiral and possess mirror symmetry with respect to the tube axis, with their nomenclature reflecting the characteristic atomic arrangements at the tube edge. 
The phonon properties of these achiral CNTs have been extensively investigated both theoretically and experimentally~\cite{rf:Saito02, rf:Suzuura, rf:Mahan, rf:Wang, rf:YamamotoPRL2004, rf:YamamotoAPE2009, rf:YamamotoPRL2006}. 
In contrast, chiral CNTs lack such symmetry and are intrinsically nonsymmetric. They exhibit handedness and are therefore not superimposable on their mirror images. 
As a consequence of this structural asymmetry, chiral CNTs are expected to host intrinsic chiral phonons and to exhibit thermally induced EdH effects.

In this Letter, we present a comprehensive theoretical study of phonon angular momentum in semiconducting chiral single-walled carbon nanotubes (SWCNTs), 
focusing on the dependence of thermally induced phonon angular momentum on the tube diameter and chiral angle within the framework of the Boltzmann transport 
equation. Furthermore, we estimate the angular velocity of chiral SWCNTs under a temperature gradient arising from the thermally induced EdH effect.


First, we briefly describe a general formulation for calculating phonon angular momentum~\cite{rf:Hamada, rf:Ohe, rf:Chen, rf:Zhang01} and for quantifying the thermally 
induced EdH effect. For band insulators, where the charge and spin degrees of freedom of electrons can be neglected, the total angular momentum of 
the system under a temperature gradient can be expressed solely in terms of atomic degrees of freedom as
\begin{eqnarray}
\bm{L}_{\rm tot} = \bm{L}_{\rm lat} + \bm{L}_{\rm ph},
\label{eq:total_momentum}
\end{eqnarray}
where $\bm{L}_{\rm lat}$ denotes the angular momentum associated with rigid-body motion, and $\bm{L}_{\rm ph}$ is the angular momentum carried by phonons. 
The phonon angular momentum $\bm{L}_{\rm ph}$ is given by~\cite{rf:Ohe, rf:Chen, rf:Zhang01}
\begin{eqnarray}
\bm{L}_{\rm ph} = \sum_{\bm{q}\lambda} \bm{l}_{\bm{q}\lambda}
\left\{ n(\omega_{\bm{q}\lambda}) + \frac{1}{2} \right\},
\label{eq:OAM}
\end{eqnarray}
where $n(\omega_{\bm{q}\lambda}) = n_{0}(\omega_{\bm{q}\lambda}) + \delta n$, and
$n_{0}(\omega_{\bm{q}\lambda}) = 1/(e^{\hbar\omega_{\bm{q}\lambda}/k_{\rm B}T}-1)$
is the Bose--Einstein distribution function for the phonon branch $\lambda$ with wave vector $\bm{q}$ at temperature $T$. Here, $\delta n$ represents the deviation from 
equilibrium induced by the temperature gradient, and $\bm{l}_{\bm{q}\lambda}$ is the angular momentum of a phonon mode characterized by 
$\bm{q}$ and $\lambda$, defined as
\begin{eqnarray}
\bm{l}_{\bm{q}\lambda} = i\hbar \sum_{j}
\bm{e}_{\bm{q}j\lambda} \times \bm{e}^{\ast}_{\bm{q}j\lambda},
\label{eq:PAM}
\end{eqnarray}
where $\bm{e}_{\bm{q}j\lambda}$ is the normalized displacement eigenvector of the phonon mode $(\bm{q},\lambda)$ corresponding to the $j$th atom in the unit cell. 
The phonon eigenvectors are obtained by solving the eigenvalue problem
\begin{eqnarray}
\omega_{\bm{q}\lambda}^{2} e^{\alpha}_{\bm{q}j\lambda}
- \sum_{j'} \sum_{\beta}
D^{\beta j'}_{\alpha j}(\bm{q})\, e^{\beta}_{\bm{q}j'\lambda} = 0,
\label{eq:DM}
\end{eqnarray}
where $D^{\beta j'}_{\alpha j}(\bm{q})$ is the dynamical matrix.

\begin{figure}[t]
\begin{center}
\includegraphics[width=70mm]{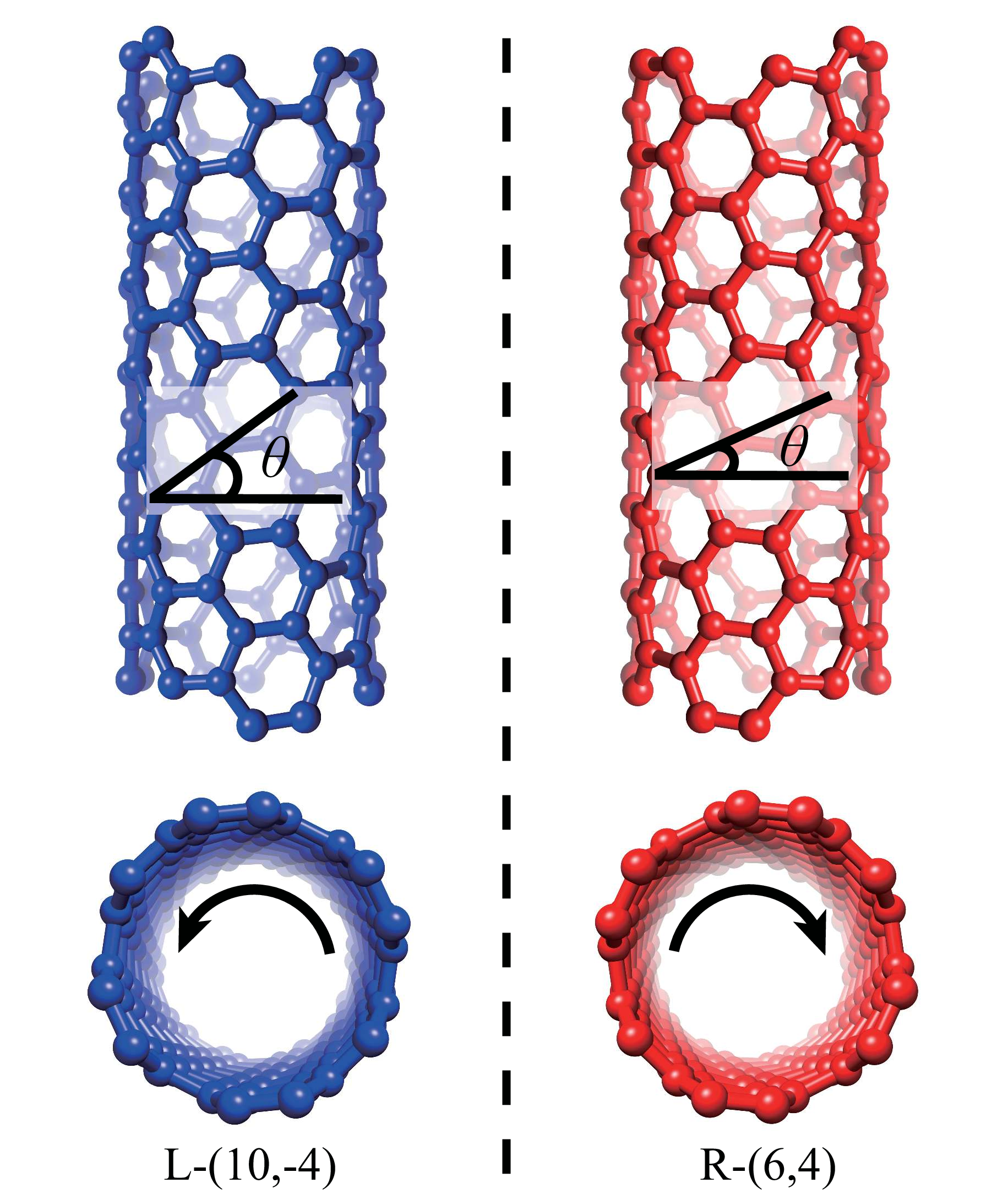}
\caption{(Color online)
Illustrations of the unit-cell structures of left- (L-) and right-handed (R-) chiral (10,$-4$) and (6,4) SWCNTs, shown in blue and red, respectively. 
The chiral angle $\theta$ lies in the ranges $30^\circ < \theta < 60^\circ$ for L-handed structures and $0^\circ < \theta < 30^\circ$ for R-handed structures.
}
\label{fig:01}
\end{center}
\end{figure}

In equilibrium, the populations of two enantiomeric chiral phonons with wave vectors $\bm{q}$ and $-\bm{q}$ are identical ($\delta n = 0$), resulting in 
a vanishing phonon angular momentum, $\bm{L}_{\rm ph}^{\rm eq} = 0$, where $\bm{L}_{\rm ph}^{\rm eq}$ denotes the phonon angular momentum in equilibrium. 
In contrast, when a temperature gradient is applied to the phonon system along the $\beta$ direction, an imbalance between the populations of the two enantiomeric 
chiral phonons is induced ($\delta n \neq 0$), leading to a finite net phonon angular momentum $\Delta L_{\rm ph}^{\alpha}$. 

In this situation, the nonequilibrium distribution $\delta n$ can be described by the Boltzmann equation within the relaxation-time approximation as
\begin{eqnarray}
\delta n = \frac{1}{T}\sum_{\beta}\hbar\omega_{\bm{q}\lambda}
v^{\beta}_{\bm{q}\lambda}\tau_{\bm{q}\lambda}
\left( -\frac{\partial n_{0}}{\partial \hbar\omega_{\bm{q}\lambda}} \right)
(-\nabla_{\beta} T),
\label{eq:delta_n}
\end{eqnarray}
where $v_{\bm{q}\lambda}$ is the phonon group velocity and $\tau_{\bm{q}\lambda}$ is the phonon relaxation time. 
Substituting Eq.~(\ref{eq:delta_n}) into Eq.~(\ref{eq:OAM}), the phonon angular momentum can be expressed as
$L_{\rm ph}^{\alpha} = L_{\rm ph}^{\rm eq,\alpha} + \Delta L_{\rm ph}^{\alpha}$, where $\Delta L_{\rm ph}^{\alpha}$ is given by
\begin{eqnarray}
\Delta L_{\rm ph}^{\alpha}
= \frac{1}{T}\sum_{\beta}\sum_{\bm{q}\lambda}
\hbar\omega_{\bm{q}\lambda}
l^{\alpha}_{\bm{q}\lambda}
v^{\beta}_{\bm{q}\lambda}
\tau_{\bm{q}\lambda}
\left( -\frac{\partial n_{0}}{\partial \hbar\omega_{\bm{q}\lambda}} \right)
(-\nabla_{\beta} T).
\label{eq:OAM_temperature}
\end{eqnarray}

\begin{figure}[t]
\begin{center}
\includegraphics[width=80mm]{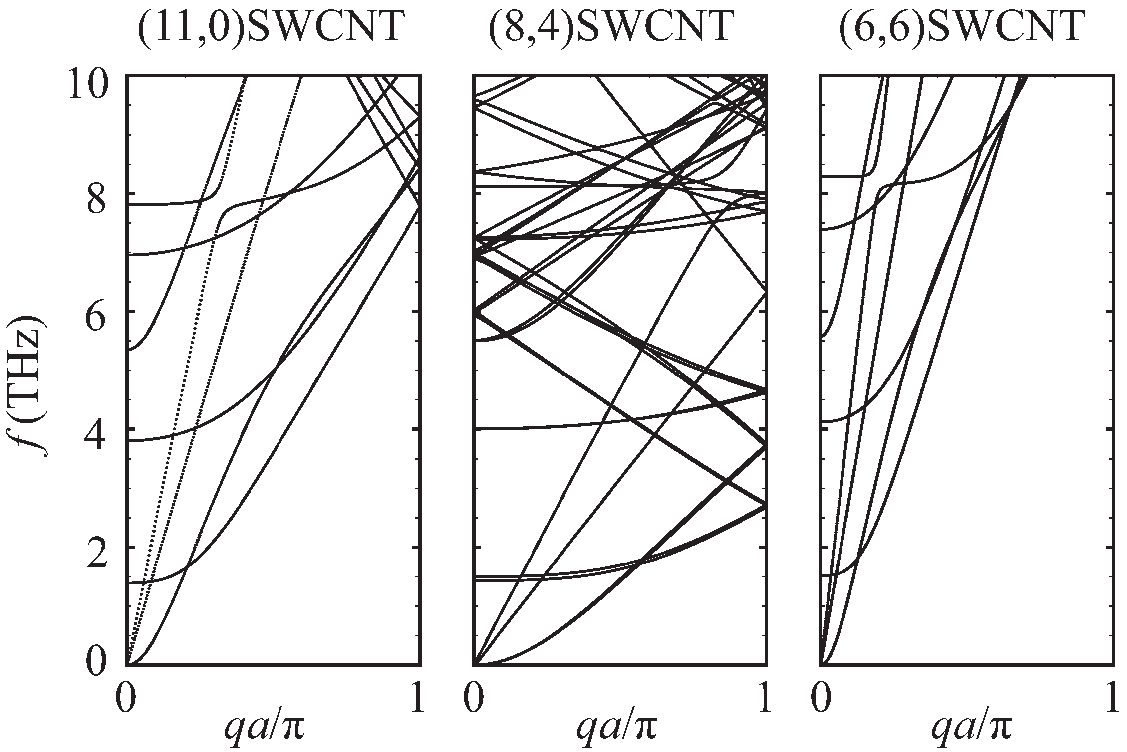}
\caption{
Phonon dispersion relations of (11,0), (8,4), and (6,6) SWCNTs with unit-cell lengths $a = 0.43$, $1.15$, and $0.25$~nm, respectively. 
}
\label{fig:02}
\end{center}
\end{figure}

Accordingly, the phonon angular momentum per unit volume induced by a temperature gradient along the $\beta$ direction is defined as~\cite{rf:Hamada, rf:Ohe, rf:Chen}
\begin{eqnarray}
J_{\rm ph}^{\alpha}
= \frac{\Delta L_{\rm ph}^{\alpha}}{V}
\equiv \sum_{\beta}\kappa^{\alpha\beta} (-\nabla_{\beta} T),
\label{eq:total_PAM}
\end{eqnarray}
where $V$ denotes the system volume and $\kappa^{\alpha\beta}$ is the thermal angular momentum coefficient, given by
\begin{eqnarray}
\kappa^{\alpha\beta}
= \frac{1}{VT}\sum_{\bm{q}\lambda}
\hbar\omega_{\bm{q}\lambda}
l^{\alpha}_{\bm{q}\lambda}
v^{\beta}_{\bm{q}\lambda}
\tau_{\bm{q}\lambda}
\left( -\frac{\partial n_{0}}{\partial \hbar\omega_{\bm{q}\lambda}} \right).
\label{eq:thermal_transport_coefficient}
\end{eqnarray}

Based on the conservation of total angular momentum described in Eq.~(\ref{eq:total_momentum}), we obtain
\begin{eqnarray}
\Delta L_{\rm lat}^{\alpha} = - \Delta L_{\rm ph}^{\alpha}.
\label{eq:conservation_momentum}
\end{eqnarray}
This relation implies that the system undergoes rigid-body rotation about the $\alpha$ axis, which is referred to as the thermally induced EdH effect. 
By evaluating the thermal angular momentum coefficient $\kappa^{\alpha\beta}$ in Eq.~(\ref{eq:thermal_transport_coefficient}), the angular momentum associated with rigid-body rotation, $\Delta L_{\rm lat}^{\alpha}$, can be quantitatively determined.


In this Letter, we focus on semiconducting chiral $(n,m)$ SWCNTs with various tube diameters $d_{\rm t}$ and chiral angles 
$\theta$. Figure~\ref{fig:01} illustrates the left- (L-) and right-handed (R-) unit-cell structures of chiral (10,$-4$) and (6,4) SWCNTs. The chiral angle $\theta$ is 
defined as shown in Fig.~\ref{fig:01}; for L- and R-handed structures, $\theta$ lies in the ranges $30^\circ < \theta < 60^\circ$ and $0^\circ < \theta < 30^\circ$, 
respectively. The limiting cases $\theta = 0^\circ$ and $30^\circ$ correspond to the highly symmetric zigzag and armchair SWCNTs, respectively. 

We define the tube axis along the $x$ direction and the system volume as $V = AL$, where $L$ is the tube length and $A = \pi d_{\rm t} t$ is the effective cross-sectional area, 
with the tube thickness taken as $t = 3.4$~\AA~\cite{rf:Pop, rf:Gu}.

\begin{figure}[t]
\begin{center}
\includegraphics[width=80mm]{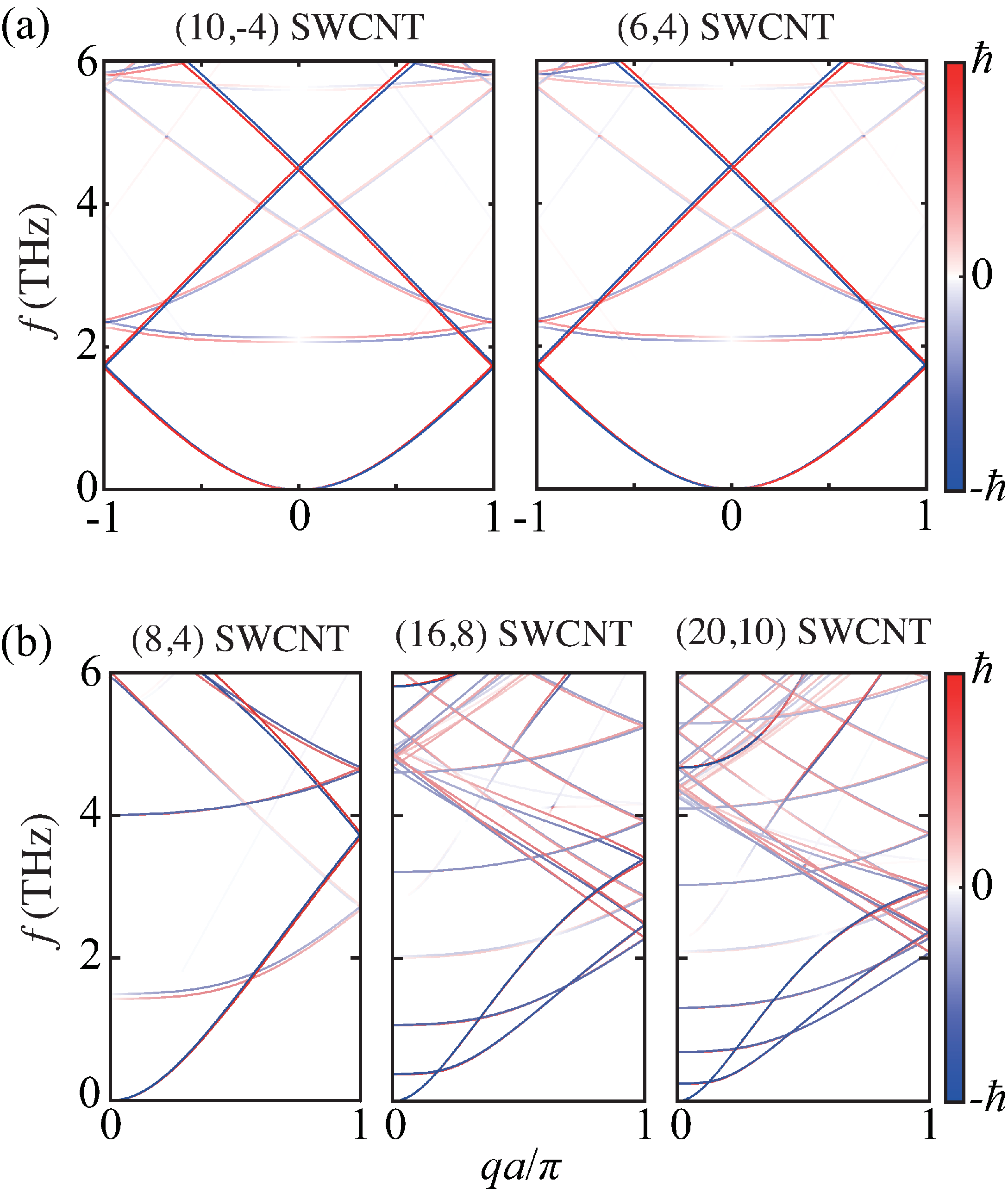}
\caption{(Color online)
(a) Phonon angular momentum $l_{q\lambda}^{x}$ for (10,$-4$) and (6,4) SWCNTs with unit-cell length $a = 1.89$~nm.
The blue and red colors represent the amplitudes of phonon angular momentum for left- and right-handed chiral phonon modes, respectively. 
(b) $l_{q\lambda}^{x}$ for (8,4), (16,8), and (20,10) SWCNTs with a fixed chiral angle $\theta = 19.1^\circ$ and unit-cell length $a = 1.14$~nm. 
The corresponding tube diameters are $d_{\rm t} = 0.85$, $1.68$, and $2.10$~nm, respectively.
}
\label{fig:03}
\end{center}
\end{figure}


Figure~\ref{fig:02} shows the phonon dispersion relations of (11,0), (8,4), and (6,6) SWCNTs. The phonon dispersions and eigenvectors are calculated from the dynamical 
matrix defined in Eq.~(\ref{eq:DM}) for optimized atomic structures. The interatomic forces between carbon atoms are modeled using the Tersoff potential~\cite{rf:lindsay}, 
which is known to provide a reliable description of carbon--carbon interactions in SWCNTs. Among the four acoustic phonon modes in SWCNTs, the two modes exhibiting 
a quadratic dependence on the one-dimensional wave vector $q$ correspond to the transverse acoustic (TA) phonon modes~\cite{rf:Suzuura}. As shown in Fig.~\ref{fig:02}, the TA phonon 
modes as well as some optical phonon modes are doubly degenerate in the highly symmetric zigzag and armchair SWCNTs.

Figure~\ref{fig:03}(a) shows the phonon angular momentum $l_{q\lambda}^{x}$ for enantiomeric L- and R-handed structures of (10,$-4$) and (6,4) SWCNTs, 
calculated using Eq.~(\ref{eq:PAM}). The blue and red colors represent the amplitudes of phonon angular momentum for L- and R-handed chiral phonon modes, respectively. 
In chiral SWCNTs, the TA phonon modes and certain optical phonon modes split into two branches carrying opposite signs of phonon angular momentum, as indicated by 
the red and blue colors. In addition, zone folding occurs at the Brillouin-zone boundary for these split branches, as shown in Fig.~\ref{fig:03}(a). These features are characteristic 
signatures of chiral SWCNTs. Furthermore, the L- and R-handed structures exhibit phonon angular momenta of opposite sign, reflecting their enantiomeric nature.

Figure~\ref{fig:03}(b) shows $l_{q\lambda}^{x}$ for R-handed (8,4), (16,8), and (20,10) SWCNTs at a fixed chiral angle $\theta = 19.1^\circ$, with corresponding tube diameters 
$d_{\rm t} = 0.85$, $1.68$, and $2.10$~nm, respectively. With increasing $d_{\rm t}$, the optical phonon modes shift toward lower frequencies. Focusing on relatively low-frequency 
chiral phonon modes, the splitting between the phonon branches is found to gradually decrease as $d_{\rm t}$ increases, indicating a tendency for the L- and R-handed chiral 
phonon branches to approach each other and eventually converge.

\begin{figure}[t]
\begin{center}
\includegraphics[width=80mm]{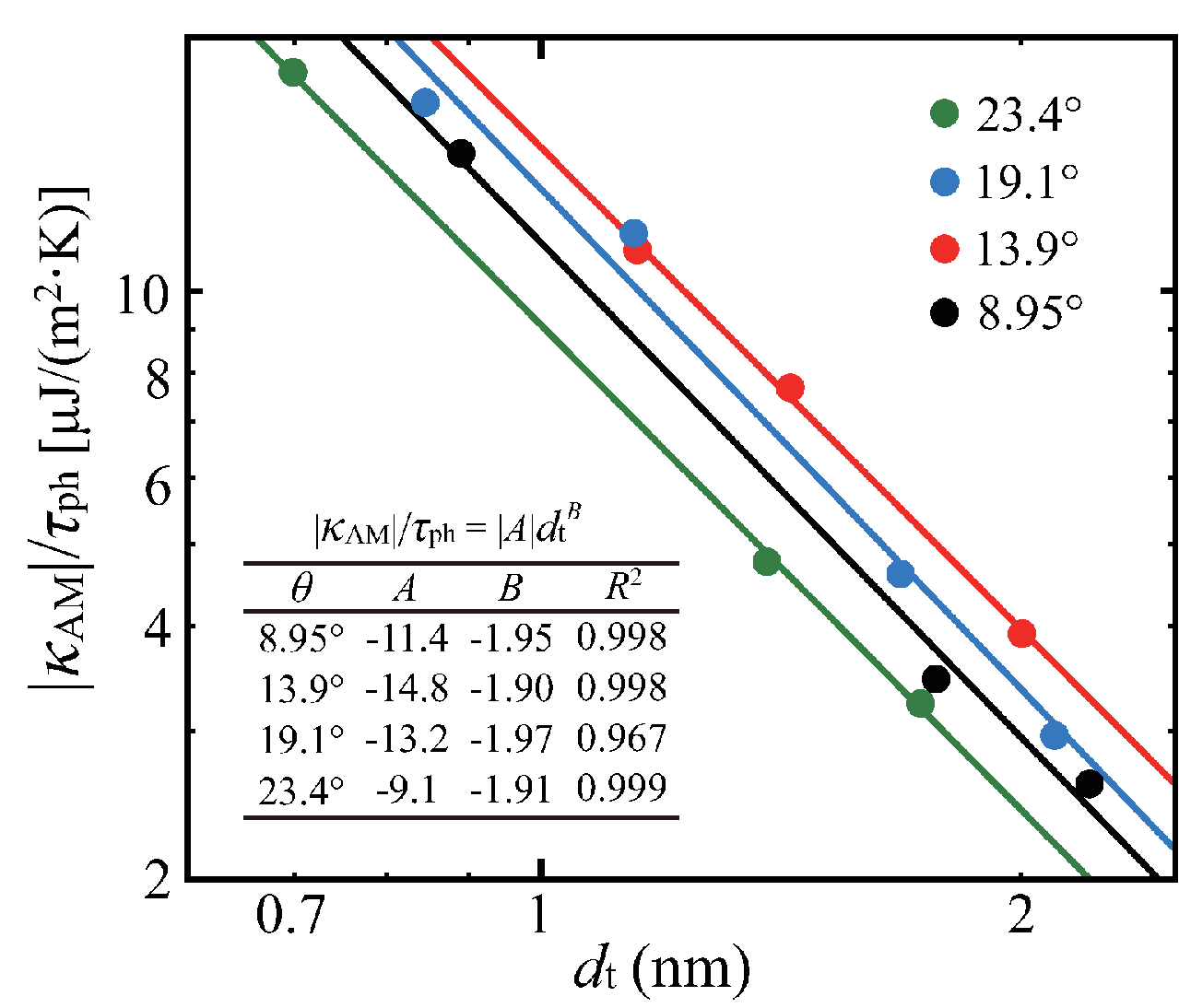}
\caption{(Color online)
Dependence of $|\kappa_{\rm AM}|/\tau_{\rm ph}$ on the tube diameter $d_{\rm t}$ for chiral SWCNTs with four different chiral angles $\theta$ at $300$~K. 
The corresponding values of $\theta$ are $8.95^\circ$, $13.9^\circ$, $19.1^\circ$, and $23.4^\circ$. The solid curves represent power-law fits of the form 
$|\kappa_{\rm AM}|/\tau_{\rm ph} = |A| d_{\rm t}^{B}$, where $A$ and $B$ are fitting parameters.
}
\label{fig:04}
\end{center}
\end{figure}

In this section, we discuss the dependence of the thermal angular momentum coefficient on the tube diameter $d_{\rm t}$ for chiral SWCNTs, 
focusing in particular on R-handed structures. 
From the symmetry of the SWCNT structure, $\kappa^{\alpha\beta}$ has finite values only in its diagonal components~\cite{rf:Hamada}.
Furthermore, taking into account the one-dimensional nature of SWCNTs, we have $v^{y}_{q\lambda} = v^{z}_{q\lambda} = 0$, which implies that $\kappa^{yy} = \kappa^{zz} = 0$. Therefore, we focus on $\kappa^{xx}(\equiv \kappa_{\rm AM})$. The coefficient $\kappa_{\rm AM}$ can be evaluated from Eq.~(\ref{eq:thermal_transport_coefficient}) under a temperature gradient applied along the tube axis. The phonon relaxation time $\tau_{q\lambda}$ generally depends on phonon modes and frequencies and exhibits complex behavior in SWCNTs~\cite{rf:Gu, rf:Cao}. However, in order to clarify the dependences on $d_{\rm t}$ and the chiral angle $\theta$ in this Letter, we assume a constant phonon relaxation time $\tau_{\rm ph}$ for simplicity.

Figure~\ref{fig:04} shows the dependence of $|\kappa_{\rm AM}|/\tau_{\rm ph}$ on $d_{\rm t}$ for chiral SWCNTs with four different chiral angles $\theta$ at $300$~K, 
namely $\theta = 8.95^\circ$, $13.9^\circ$, $19.1^\circ$, and $23.4^\circ$. The solid curves represent power-law fits of the form
$|\kappa_{\rm AM}|/\tau_{\rm ph} = |A| d_{\rm t}^{B}$, where $A$ and $B$ are the fitting parameters.
The fitting parameters and the coefficient of determination for the overall fitted curves $R^{2}$ are summarized in the inset of Fig.~\ref{fig:04}. We find that $|\kappa_{\rm AM}|/\tau_{\rm ph}$ approximately follows 
a $d_{\rm t}^{-2}$ dependence for all values of $\theta$. This result indicates that the generation of phonon angular momentum induced by chiral phonons is 
more efficient in SWCNTs with smaller diameters. This trend can be understood from the fact that the two transverse acoustic modes and the relevant optical modes 
approach each other as $d_{\rm t}$ increases, as shown in Fig.~\ref{fig:03}(b).

\begin{figure}[t]
\begin{center}
\includegraphics[width=80mm]{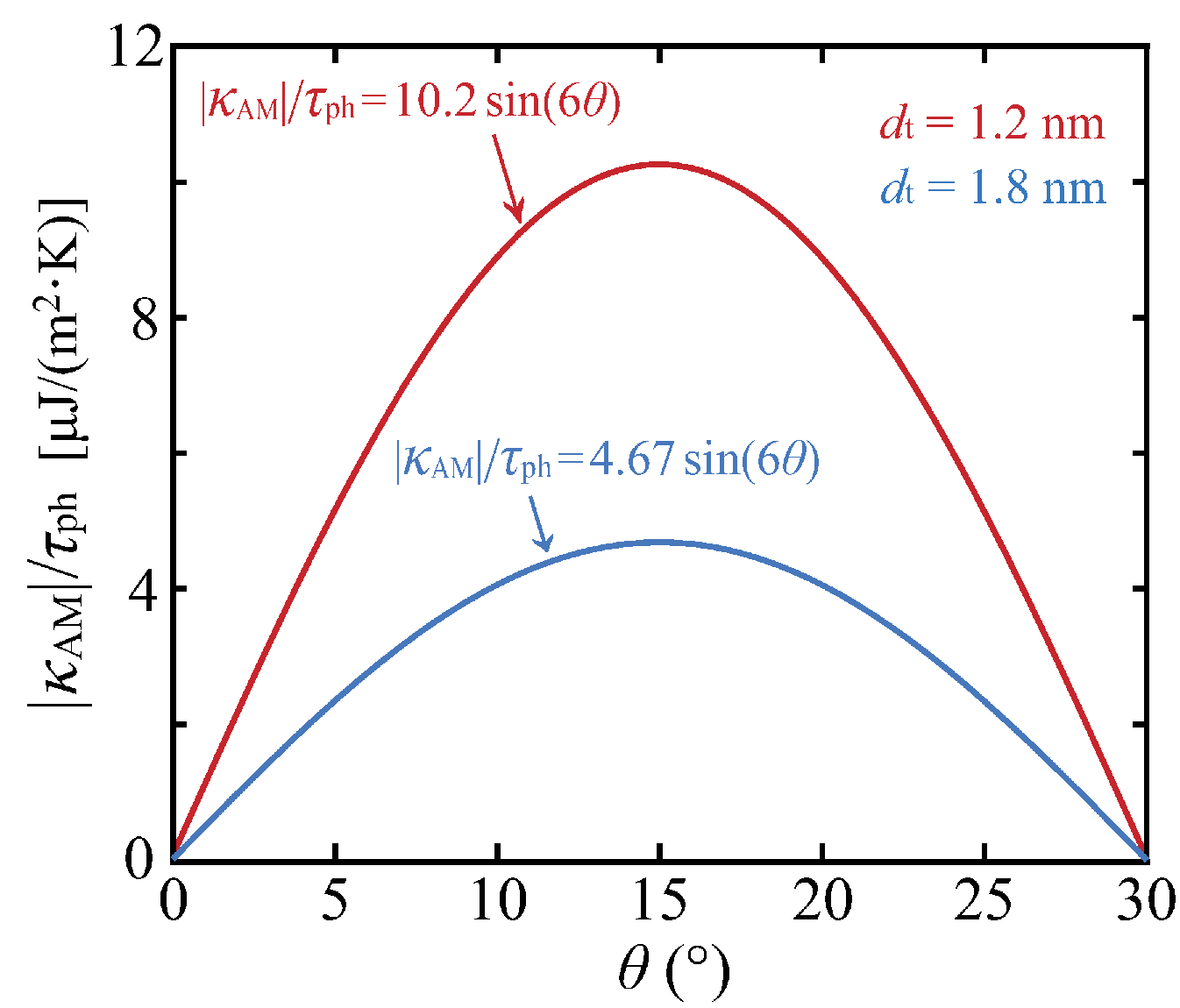}
\caption{(Color online)
Dependence of $|\kappa_{\rm AM}|/\tau_{\rm ph}$ on the chiral angle $\theta$ for chiral SWCNTs at $300$~K. The corresponding tube diameters are $d_{\rm t} = 1.2$ 
and $1.8$~nm. The red and blue curves represent sinusoidal fits of the form $|\kappa_{\rm AM}|/\tau_{\rm ph} = |C|\sin(D\theta)$, where $C$ and $D$ are fitting parameters. 
The fitting parameters $(C,D)$ are $(-10.2,\,6.0)$ and $(-4.67,\,6.0)$, respectively.
}
\label{fig:05}
\end{center}
\end{figure}

To examine the chiral-angle dependence of $|\kappa_{\rm AM}|/\tau_{\rm ph}$, we fix the tube diameter $d_{\rm t}$ and use the fitting parameters obtained in Fig.~\ref{fig:04}. 
Figure~\ref{fig:05} shows the $\theta$ dependence of $|\kappa_{\rm AM}|/\tau_{\rm ph}$ for R-handed chiral SWCNTs with fixed diameters of $d_{\rm t} = 1.2$ and 
$1.8$~nm at $300$~K. The red and blue curves correspond to sinusoidal fits of the form
$|\kappa_{\rm AM}|/\tau_{\rm ph} = |C| \sin(D\theta)$,
where the fitting parameters $C$ and $D$ are given in the inset of Fig.~\ref{fig:05}. 
We find that $|\kappa_{\rm AM}|/\tau_{\rm ph}$ reaches its maximum at $\theta = 15^\circ$, indicating that $|\kappa_{\rm AM}|$ is maximized in chiral SWCNTs with strong 
helicity, corresponding to intermediate chiral angles between the zigzag ($\theta = 0^\circ$) and armchair ($\theta = 30^\circ$) limits.

Finally, we discuss the angular velocity of chiral SWCNTs under a temperature gradient based on the thermally induced EdH effect. The energy band gap of typical intrinsic semiconducting SWCNTs is approximately 1~eV; thus, thermal excitation of electrons is negligible at room temperature. Although electron-phonon interactions persist at this temperature, their impact on the spin-orbit interaction is considered minimal due to the strong covalent bonding of carbon atoms~\cite{rf:Ando}. Therefore, we assumed that the contribution of electron spin to the angular momentum is negligible. To estimate the angular 
velocity in this Letter, we treat a chiral SWCNT as a rigid body. As is well known from classical mechanics, the angular momentum associated with rigid-body rotation satisfies
$J_{\rm ph}^{x} V = - I \omega$~\cite{rf:Chen}, in accordance with the conservation of total angular momentum described by Eq.~(\ref{eq:conservation_momentum}). 
Here, $I = M\{(r -t/2)^{2}+(r+t/2)^{2}\}/2$ is the moment of inertia, where $M$ is the total mass of the nanotube and $r$ is its radius. Accordingly, the angular velocity of rigid-body rotation in 
a chiral SWCNT is given by
\begin{eqnarray}
\omega = \frac{- J_{\rm ph}^{x} A L}{I}.
\label{eq:angular_velocity}
\end{eqnarray}

Using realistic parameters, $\nabla_{x}T = 10$~K, $\tau_{\rm ph} = 10$~ps, and $L = 1.24~\mu$m, we estimate the angular velocity for a (6,5) SWCNT with $d_{\rm t} = 0.76$~nm 
and $\theta = 27^\circ$. At $300$~K, using $\kappa_{\rm AM}/\tau_{\rm ph} = -11.21~\mu{\rm J}/({\rm m}^{2}\cdot{\rm K})$, we obtain $\omega \sim 1$~rad/s. 
This estimated angular velocity is significantly larger than previously reported values for chiral crystals~\cite{rf:Hamada, rf:Chen}. 
The predicted large angular velocity is attributed to the repeated folding of phonon branches with high group velocities, including the TA mode, at the Brillouin zone boundaries. More fundamentally, the critical factor, which is intrinsically linked to the origin of this zone folding, is the relatively small moment of inertia of the SWCNT structure compared to that of other chiral crystals, resulting from its one-dimensional cylindrical geometry.

{\bf{Conclusion}}

In this Letter, we have investigated the phonon angular momentum $l_{q\lambda}^{x}$ and the dependences of the thermal angular momentum coefficient $\kappa_{\rm AM}$ on 
the tube diameter $d_{\rm t}$ and chiral angle $\theta$ at $300$~K for various semiconducting chiral SWCNTs. We have shown that the transverse acoustic modes and 
diameter-dependent optical phonon modes, which are degenerate in zigzag and armchair SWCNTs, split into two branches in chiral SWCNTs. We find that $\kappa_{\rm AM}$ for 
chiral SWCNTs approximately scales as $d_{\rm t}^{-2}$ at a fixed chiral angle $\theta$. In addition, $|\kappa_{\rm AM}|$ exhibits a $\sin(6\theta)$ dependence at a fixed tube 
diameter, reaching its maximum at $\theta = 15^\circ$. Consequently, chiral SWCNTs with smaller diameters and intermediate chiral angles between the highly symmetric zigzag 
and armchair limits generate the largest phonon angular momentum. Furthermore, we have demonstrated that even a (6,5) SWCNT with $\theta = 27^\circ$, which is close to 
the armchair configuration, exhibits a larger angular velocity under a temperature gradient than previously studied chiral crystals. These results indicate that the thermally induced 
EdH effect in chiral SWCNTs is potentially observable in experiments.

\vspace{0.5cm}

\begin{acknowledgments}
We are grateful to M. Ogata, T. Suenaga and A. Iida for the discussion. This work is supported by Grants-in-Aid for Scientific Research from the Japan Society for the Promotion of 
Science (No. 22KK0228 and No. 25K07199).
\end{acknowledgments}

\vspace{0.5cm}

\noindent * takahiro@rs.tus.ac.jp

\end{document}